# THERMAL ANALYSIS & OPTIMIZATION OF A 3 DIMENSIONAL HETEROGENEOUS STRUCTURE

Ramya Menon C.[1] and Vinod Pangracious[2]

[1]Department of Electronics & Communication Engineering, Sahrdaya College of Engineering & Technology, Kerala, India

ramyamenonc@gmail.com

[2]Vinod Pangracious, University of Pierre and Marie Curie, Paris, France

pangracious@googlemail.com

## ABSTRACT

Besides the lot of advantages offered by the 3D stacking of devices in an integrated circuit there is a chance of device damage due to rise in peak temperature value. Hence, in order to make use of all the potential benefits of the vertical stacking a thermal aware design is very essential. The first step for designing a thermal aware architecture is to analyze the hotspot temperature generated by the devices. In this paper we are presenting the results of our thermal analysis experiments of a 3D heterogeneous structure with three layers. The bottom layer had eight identical processors at 2.4 GHz and the top layer was with four memory units. The intermediate layer was a thermal interface material (TIM). The 2D thermal analysis of the top and bottom layers was also done separately. In the next step simulations were carried out by varying TIM thickness and conductivity to study its affect on hotspot temperature so as to optimize the temperature distribution.

## 1. INTRODUCTION

Three dimensional integration of circuit is a developing technology which is gaining importance day by day. Various experiments have been done with homogeneous architecture with DRAM to increase the memory capacity. To improve performance by reducing the wire length, separating memory and processor and arranging them in different layers is a preferred design scenario. We have done experiments with heterogeneous structure involving processors and memory units to analyze the hotspot temperature by powering the devices in different ways. In our experiments we used epoxy based resin as the thermal interface material. The thickness and conductivity of TIM was varied to study the effect on hotspot temperature. Conductivity of TIM can be varied by doping with copper. The simulator used is Hotspot an efficient tool that is widely used for thermal analysis.

## 2. PROBLEM FORMULATION

A three layered heterogeneous structure with eight processor in the bottom layer and four memory units in the topmost layer was taken. An intermediate layer of thermal interface material was placed in between these layers. The memory was L2 cache that consumed a power of 1.3W and the processor consumed a power of 50.9W. The floorplans of the processor and memory layers are as shown in fig.1. The crossbar and cb are bare Silicon areas reserved for introducing through Silicon vias (TSVs). M1, M2, M3 and M4 represent the memory units .The eight processors are represented by P1, P2, P3, P4, P5, P6, P7 and P8.

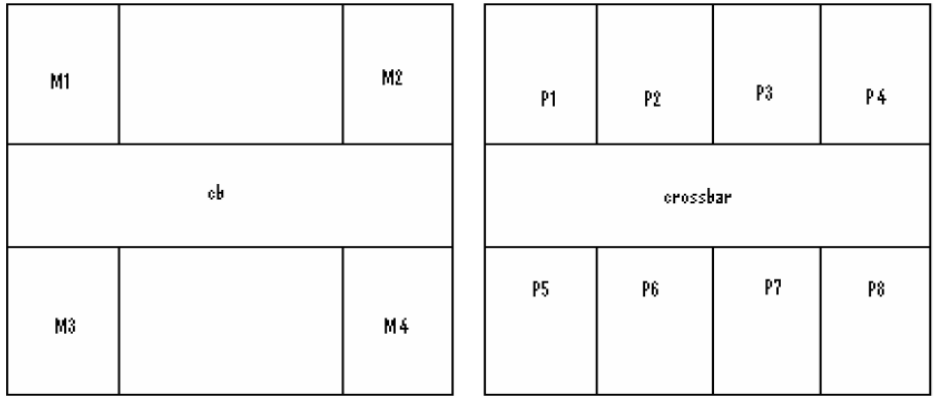

Figure1. The floorplan of memory and processor layers

## 3. TWO DIMENSIONAL SIMULATIONS & ANALYSIS

The 2D analysis of these layers was done separately to find hotspot temperature. The profiles obtained is shown in fig.2. The profile in the left was obtained while analyzing the processor layer alone and the one in the right was obtained by analyzing the memory layer alone.

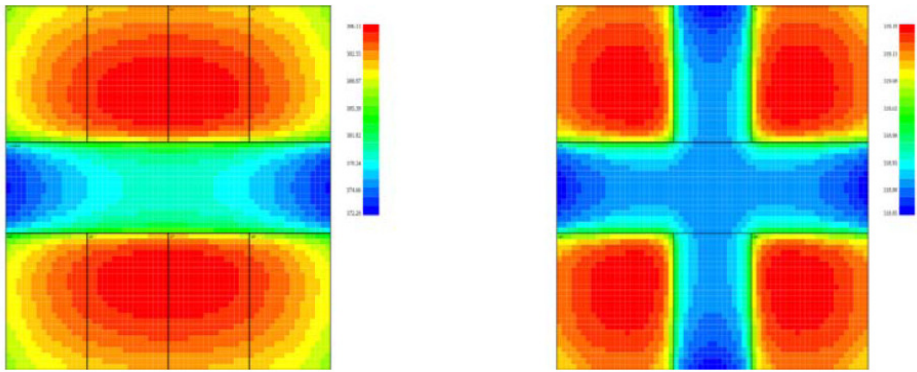

Figure 2. Two dimensional thermal profiles

The hotspot was developed P2, P3, P6 & P7 of the processor layer and all the memory units in the memory layer. The hotspot temperature in the processor layer was 396.13K and in the memory layer was 319K. The temperature generated by processors is very high when compared to memory layer.

## 4. THREE DIMENSIONAL SIMULATIONS & ANALYSIS

A 3D structure was formed with three layers. The bottom layer (layer 1) was the processor layer and the top layer (layer 3) was the memory layer. The intermediate layer (layer 2) was TIM. A total of five simulations were carried out. In the first simulation only processor layer was powered. No power supply connection was given to memory layer. This simulation was done to study the affect of processors alone on hotspot temperature. However, in real case a processor cannot work without powering on the cache memory. Only memory layer was powered in the second simulation. Both the layers were simultaneously powered in the third simulation. In the fourth simulation, the memory layer was powered and only four processors at the corners i.e. P1, P4, P5 & P8 in the bottom layer were powered. Finally, in the fifth simulation, the memory layer was powered and four processors - P2, P3, P6 & P7 in the bottom layer were powered. The peak temperature developed in each of the units in five simulations is shown in tables below. The first and third rows show the unit names (U) and the second and fourth rows show the temperature (T) developed in each of the units in kelvin. The crossbar in the processor layer is represented by Cb1. In all these simulations TIM thickness was $2x10^{-5}$m.

Table 1. Only processor layer powered

| U | P1 | P2 | P3 | P4 | Cb1 | P5 | P6 |
|------|--------|--------|--------|--------|--------|--------|--------|
| T(K) | 395.41 | 398.51 | 398.51 | 395.41 | 378.85 | 395.41 | 398.51 |
| U | P7 | P8 | M1 | M2 | M3 | M4 | cb |
| T(K) | 398.51 | 395.41 | 382.86 | 382.86 | 382.86 | 382.86 | 378.73 |

Table 2. Only memory layer powered

| U | P1 | P2 | P3 | P4 | Cb1 | P5 | P6 |
|------|--------|--------|--------|--------|--------|--------|--------|
| T(K) | 319.00 | 318.95 | 318.95 | 319.00 | 318.90 | 319.00 | 318.95 |
| U | P7 | P8 | M1 | M2 | M3 | M4 | cb |
| T(K) | 318.95 | 319.00 | 319.01 | 319.01 | 319.01 | 319.01 | 318.90 |

Table 3. Both layers powered

| U | P1 | P2 | P3 | P4 | Cb1 | P5 | P6 |
|------|--------|--------|--------|--------|--------|--------|--------|
| T(K) | 396.26 | 399.30 | 399.30 | 396.26 | 379.60 | 396.26 | 399.30 |
| U | P7 | P8 | M1 | M2 | M3 | M4 | cb |
| T(K) | 399.30 | 396.26 | 383.72 | 383.72 | 383.72 | 383.72 | 379.48 |

Table 4. Hotspot temperature values obtained in the fourth simulation

| U | P1 | P2 | P3 | P4 | Cb1 | P5 | P6 |
|------|--------|--------|--------|--------|--------|--------|--------|
| T(K) | 368.23 | 347.03 | 347.03 | 368.23 | 347.09 | 368.23 | 347.03 |
| U | P7 | P8 | M1 | M2 | M3 | M4 | cb |
| T(K) | 347.03 | 368.23 | 354.04 | 354.04 | 354.04 | 354.04 | 347.08 |

Table 5. Hotspot temperature values obtained in the fifth simulation

| U | P1 | P2 | P3 | P4 | Cb1 | P5 | P6 |
|------|--------|--------|--------|--------|--------|--------|--------|
| T(K) | 347.03 | 371.22 | 371.22 | 347.03 | 351.41 | 347.03 | 371.22 |
| U | P7 | P8 | M1 | M2 | M3 | M4 | cb |
| T(K) | 371.22 | 347.03 | 348.69 | 348.69 | 348.69 | 348.69 | 351.29 |

The hotspot temperature remained as 319K in the 2D memory experiment and in the 3D experiment with only memory layer powered. But in the case of 3D experiment with only processor layer powered there was a rise of 2.38K when compared to 2D case. When both the layers were powered the hotspot temperature again rose by 0.79K. Simulations 4 & 5 had different hotspot temperature values although both the simulations had four powered processors.

## 5. SIMULATIONS WITH VARYING TIM THICKNESS

In modern 3D technology, wafers are thinned for 3D stacking and this can cause easy damage of the device. Here with this simulation instead of decreasing wafer thickness, we varied TIM thickness to study its affect on hotspot temperature. Three simulations were carried out by varying TIM thickness. In all the simulations both the layers were powered. The thickness of the interface is $2 \times 10^{-5}$m by default in Hotspot. This value was changed by modifying the layer configuration file. With the default thickness the peak temperature values obtained in each of the units is shown in table 3. Next, the thickness of TIM is increased 10 times the default value. Now the interface layer thickness is $2 \times 10^{-4}$m .The temperature values obtained in this case is as shown in the table 6.

Table 6. Hotspot temperature with 10 times increase in TIM thickness

| U | P1 | P2 | P3 | P4 | Cb1 | P5 | P6 |
|------|--------|--------|--------|--------|--------|--------|--------|
| T(K) | 489.10 | 492.00 | 492.00 | 489.10 | 393.15 | 489.10 | 492.00 |
| U | P7 | P8 | M1 | M2 | M3 | M4 | cb |
| T(K) | 492.00 | 489.10 | 383.51 | 383.51 | 383.51 | 383.51 | 380.55 |

In the next simulation the thickness was decreased 10 times the default value. The new interface thickness was $2 \times 10^{-6}$m. The result is shown in table 7.

Table 7. Hotspot temperature with 10 times decrease in TIM thickness

| U | P1 | P2 | P3 | P4 | Cb1 | P5 | P6 |
|---|---|---|---|---|---|---|---|
| T(K) | 386.79 | 389.83 | 389.83 | 386.79 | 379.41 | 386.79 | 389.83 |
| U | P7 | P8 | M1 | M2 | M3 | M4 | cb |
| T(K) | 389.83 | 386.79 | 383.73 | 383.73 | 383.73 | 383.73 | 379.39 |

## 6. ANALYSIS OF TIM THICKNESS VARYING SIMULATIONS

Variations in TIM thickness can affect the hotspot temperature values. The hotspot temperature of various experiments is shown in the table 8.

Table 8. Hotspot temperature

| Exp. No: | Epoxy Thickness (m) | Hotspot Temperature (K) | Hotspot Location |
|---|---|---|---|
| 1 | $2 \times 10^{-5}$ | 399.30 | P2,P3,P6 & P7 |
| 2 | $2 \times 10^{-4}$ | 492.00 | P2,P3,P6 & P7 |
| 3 | $2 \times 10^{-6}$ | 389.83 | P2,P3,P6 & P7 |

The plot with epoxy thickness (m) in x-axis & hotspot temperature (K) in y-axis is shown in fig.3.

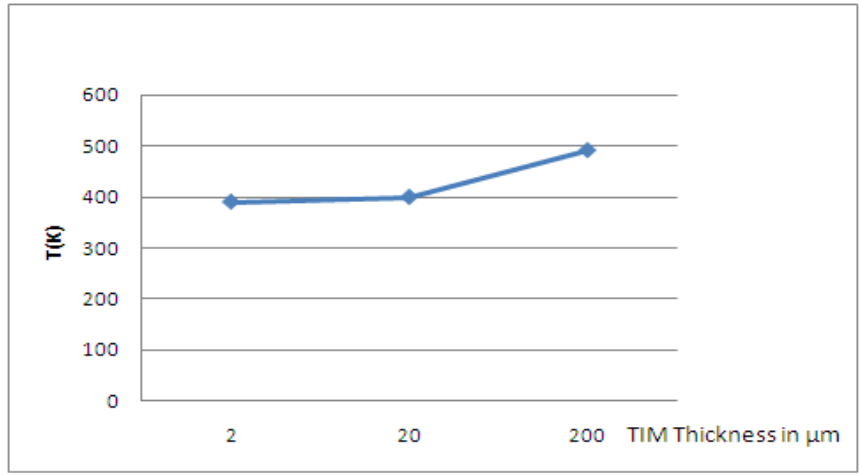

Figure 3. Temperature plot of epoxy thickness varying experiments

# 7. SIMULATIONS WITH VARYING TIM CONDUCTIVITY

TIM conductivity was varied in different simulations by copper doping. TIM used was epoxy based resin which is a dielectric material with a resistivity of 0.25mK/W.With this value the result obtained is shown in table 3. The copper was introduced uniformly throughout the epoxy based resin in varying amounts in different simulation experiments. Three simulations were carried out. In the first simulation, the TIM resistivity was made 0.125mK/W i.e. reduced to half the value of pure TIM. In the next one, this value was reduced to 0.0625mK/W. TIM resistivity was 0.1875mK/W in the third and final simulation. The hotspot temperature obtained is shown in the following tables.

Table 9. First simulation hotspot temperatures

| U | P1 | P2 | P3 | P4 | Cb1 | P5 | P6 |
|------|--------|--------|--------|--------|--------|--------|--------|
| T(K) | 391.00 | 394.04 | 394.04 | 391.00 | 379.48 | 391.00 | 394.04 |
| U | P7 | P8 | M1 | M2 | M3 | M4 | cb |
| T(K) | 394.04 | 391.00 | 383.73 | 383.73 | 383.73 | 383.73 | 379.43 |

Table 10. Second simulation hotspot temperatures

| U | P1 | P2 | P3 | P4 | Cb1 | P5 | P6 |
|------|--------|--------|--------|--------|--------|--------|--------|
| T(K) | 388.36 | 391.41 | 391.41 | 388.36 | 379.43 | 388.36 | 391.41 |
| U | P7 | P8 | M1 | M2 | M3 | M4 | cb |
| T(K) | 391.41 | 388.36 | 383.73 | 383.73 | 383.73 | 383.73 | 379.40 |

Table 11. Third simulation hotspot temperatures

| U | P1 | P2 | P3 | P4 | Cb1 | P5 | P6 |
|------|--------|--------|--------|--------|--------|--------|--------|
| T(K) | 393.63 | 396.67 | 396.67 | 393.63 | 379.53 | 393.63 | 396.67 |
| U | P7 | P8 | M1 | M2 | M3 | M4 | cb |
| T(K) | 396.67 | 393.63 | 383.72 | 383.72 | 383.72 | 383.72 | 379.45 |

# 8. ANALYSIS OF TIM CONDUCTIVITY VARYING SIMULATIONS

Variations in TIM conductivity can also affect the hotspot temperature values. The hotspot temperature of various experiments is shown in the table 12. The corresponding plot is shown in fig.4.

Table12. Hotspot temperatures of conductivity varying simulations

| Exp. No: | TIM Resistivity(mK/W) | Hotspot Temperature (K) | Hotspot Location |
|----------|----------------------|------------------------|------------------|
| 1 | 0.125 | 394.04 | P2,P3,P6 & P7 |
| 2 | 0.0625 | 391.41 | P2,P3,P6 & P7 |
| 3 | 0.1875 | 396.67 | P2,P3,P6 & P7 |

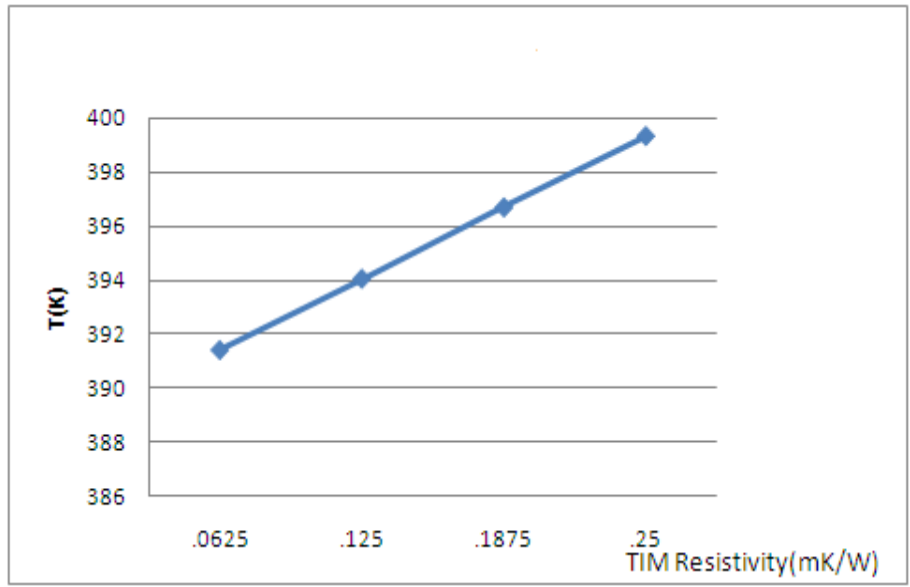

Figure 4. Temperature plot of epoxy conductivity varying experiments

From fig.4 hotspot temperature increases linearly with increase in TIM resistivity. As conductivity is the reciprocal of resistivity, the hotspot temperature decreases with increase in TIM conductivity.

## 9. CONCLUSION & FUTURE WORK

Always the bottom layers developed higher temperatures. From the first set of simulations the processor layer contributed much to rise in hotspot temperature than the memory layer. Varying the TIM thickness is a provision to transfer this temperature to upper layers and finally to heat sink and spreader. As the epoxy thickness increases there is an exponential increase in hotspot temperature. But when the thickness is reduced below certain point (2µm) the temperature value remains almost constant. So there is no use of decreasing epoxy thickness beyond a value to reduce peak temperature. When TIM resistivity increases, hotspot temperature increases linearly.TIM resistivity can be reduced by doping with copper to minimize the hotspot temperature. Our next step is to introduce Through Silicon Vias (TSVs) in the crossbar region for thermal optimization. TTSV (Thermal Through Silicon Vias) will also be introduced and studied.

**Authors**

Ramya Menon. C completed B. Tech degree in Electronics and Communication Engineering, from S.C.T College of Engineering, affiliated to Kerala University, Kerala. She did her M.Tech in VLSI and Embedded Systems, at Rajagiri School of Engineering and Technology, India. Currently working as assistant professor at Sahrdaya college of Engineering & Technology, India. The project currently working on is thermal modeling & 3D stacking of integrated circuits. She is the author of two international publications in this area.

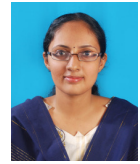

Vinod Pangracious received the B.Tech degree in Electronics Engineering from Cochin University of Kerala in 1995 and M.Tech degree from IIT Bombay India in 2000. Currently pursuing PhD at University of Pierre and Marie Curie Paris France. He is currently an Associate Professor at Rajagiri School of Engineering & Technology India. He has authored and co-authored 10 publications in these areas. He is an internationally experienced electronics engineering professional with extensive expertise in memory design, high speed digital circuit design, logic library development, verification, test and characterization. His research interest focus on design methodologies for integrated systems, including thermal management technique for multiprocessor system on chip, novel nanoscale architectures for logic and memories, dynamic memory, 3D integration and manufacturing technologies.

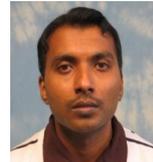